\documentclass[12pt,preprint]{aastex}
\usepackage{graphicx,amssymb,mathrsfs,amsmath,subfigure}
\begin{document}
\title{Radio emission from acceleration sites of solar flares}
\author{Yixuan Li\altaffilmark{1}, Gregory D. Fleishman\altaffilmark{1,2}}
\altaffiltext{1}{Center for Solar-Terrestrial Research, New Jersey
    Institute of Technology, Newark, NJ 07102; yl89@njit.edu, gfleishm@njit.edu}
\altaffiltext{2}{Ioffe Physico-Technical Institute, St. Petersburg
194021, Russia}

\begin{abstract}

The Letter takes up a question of what radio emission is produced by
electrons at the very acceleration site of a solar flare.
Specifically, we calculate incoherent radio emission produced within
two competing acceleration models---stochastic acceleration by
cascading MHD turbulence and regular acceleration in  collapsing
magnetic traps. Our analysis clearly demonstrates that the radio
emission from the acceleration sites: (i) has sufficiently strong
intensity to be observed by currently available radio instruments
and (ii)  has spectra and light curves which are distinctly
different in these two competing models, which makes them
observationally distinguishable. In particular, we suggest that some
of the narrowband microwave and decimeter continuum bursts may be a
signature of the stochastic acceleration in solar flares.

\end{abstract}


\section{Introduction}

Acceleration of charged particles is an internal property of energy
release in solar flares, which has not yet been fully understood in
spite of a significant progress achieved recently
\citep[e.g.,][]{Aschw_2002, Vilmer_MacKinnon_2003}. A traditional
way of getting information on the accelerated electrons in flares is
the analysis of the hard X-ray (HXR) emission produced by nonthermal
bremsstrahlung. However, because the bremsstrahlung intensity
increases with the density of the ambient plasma, it is likely that
in most cases the acceleration site and HXR emission site are
spatially separated; therefore, the HXR emission does not carry
direct information on the acceleration site. This concept of
distinct acceleration, propagation, and emission regions was then
inherited by solar radio astronomy \citep[e.g., Fig. 9 in ][]{BBG},
which looks relevant to relatively weak events of electron
acceleration visualized by coherent emission of type III groups and
of accompanying metric spikes \citep[e.g., Fig. 10 in ][]{BBG}.

However, it is well known that a charged particle produces
electromagnetic emission as it moves with acceleration. Stated
another way, fast electrons must produce radiation immediately at
the acceleration region with  intensity and other characteristics
defined by  type of the acceleration, or more precisely, by the type
of fast electron trajectories in the acceleration region. We show in
this Letter that typically this emission has a spectral peak at the
microwave range, which makes the radio observation the most suitable
to study  the acceleration region in flares.

By now, a huge variety of  acceleration mechanisms and models has
been proposed and developed for the solar flares. Acceleration by DC
electric fields, both sub-Dreicer and super-Dreicer,
\citep{Holman85, Tsuneta85, HolmanBenka, Litvinenko96,
Litvinenko_2000, Litvinenko_2003a}; stochastic acceleration by
turbulent waves \citep{Petrosian92, Petrosian94, Miller96, Miller97,
Petrosian97, Petrosian98, Byk_Fl_2009},  the classical diffusive
shock acceleration \citep{Aschw_2002}; the regular (betatron plus
Fermi) acceleration in collapsing magnetic traps
\citep{Somov_Kosugi_1997, Somov_Bogachev_2003, Karlicky_Kosugi_2004,
Bogachev_Somov_2005, Bogachev_Somov_2007, Bogachev_Somov_2009}; all
are currently considered in the context of  solar flares.

To illustrate  potential ability of  radio observations to detect
the radiation from the flare acceleration site and to distinguish
then between competing acceleration mechanisms, we calculate here
radio emission generated within two distinct acceleration
models---stochastic acceleration by a turbulence spectrum and
regular acceleration in collapsing traps. Radio emission of flares
is known to be produced by a variety of emission mechanisms
including gyrosynchrotron (GS) emission, bremsstrahlung, transition
radiation, and a number of coherent radiative processes \citep{BBG,
Nindos_etal_2008}. Some of the observed emission types can in fact
originate from acceleration sites, while others---from electrons
trapped in closed magnetic loops or from electrons propagating along
open field lines. Based on our analysis, we suggest that some of the
narrowband microwave and decimeter continuum bursts may be a
signature of the stochastic acceleration in solar flares, while the
collapsing trap acceleration must reveal itself in drifting to
higher frequencies microwave GS bursts.


\section{Radio Emission from a Region of Stochastic Acceleration} 

Basically, various models of stochastic acceleration differ from
each other by the accelerating agent (the plasma or MHD eigen-mode
responsible for the wave-particle energy exchange) and presence or
absence of some pre-acceleration (injection) process. To be
specific, we assume a 'pure' stochastic acceleration process when
the electrons are accelerated directly from the thermal pool
\citep{Petrosian92, Miller96}, perhaps as a result of MHD turbulence
cascading towards the smallest scales involved into resonant
interaction of the waves with thermal or weakly superthermal
electrons.

\begin{figure*}[!h]
\begin{center}
\epsscale{0.5} \plotone{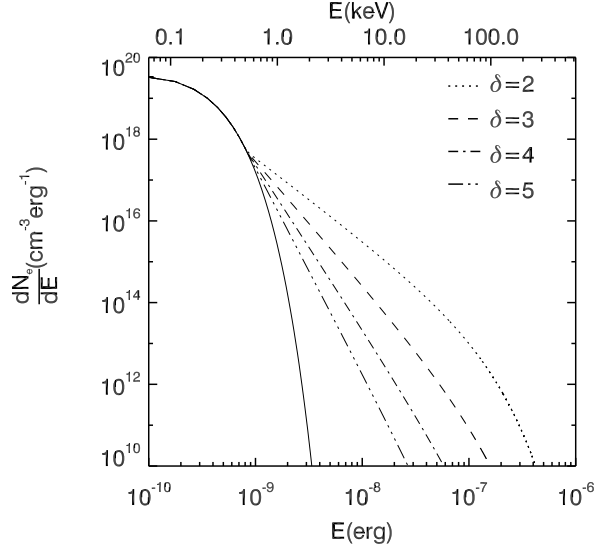} \caption{ Distribution function of
electrons for different $\delta$ and the following parameters: the
thermal electron number density $n_e= 10^{10}$ cm$^{-3}$; the
electron temperature $T_e = 10^6$ K; the matching energy is $E_0= 6
kT_e$. The solid curve denotes the original Maxwellian
distribution.} \label{FIG01}
\end{center}
\end{figure*}

Within this model the MHD turbulence is created at some large scale
and then a broad spectrum of the turbulence develops due to the
turbulence cascading. As soon as  small-scale waves capable of
resonant interaction with  electrons from  Maxwellian tail are
produced they start to accelerate those electrons. This process can
be modeled by growing a power-law tail \citep[sf., e.g. spectra of
accelerated electrons presented by][]{Petrosian92, Miller96,
Byk_Fl_2009}
\begin{equation}
\label{el_spectrum_st_acc} N(E)=(\delta(t)-1)\frac{n(>E_0)\cdot
E_0^{\delta(t)-1}}{E^{\delta(t)}}\exp\left(-\frac{E}{E_{br}(t)}\right),
\end{equation}
where the time-dependent acceleration is modeled by increasing the
break energy $E_{br}(t)$ and hardening the energy spectrum
(decreasing  spectral index $\delta(t)$). This nonthermal
distribution of accelerated electrons is assumed to match the
original Maxwellian distribution at a certain energy $E_0$;
$n(>E_0)$ is evidently defined by the matching condition:
\begin{equation}
\label{el_matching_st_acc}
n(>E_0)=\frac{2n_e}{\delta(t)-1}\sqrt{\frac{E_0^3}{\pi(kT_e)^3}}\cdot
\exp\left(\frac{-E_0}{kT_e}\right) \exp\left(\frac{E_0}{E_{br}(t)}\right), 
\end{equation}
where $n_e$ and $T_e$ are the number density and temperature of the
thermal electrons, $k$ is the Boltzman constant. Figure~\ref{FIG01}
shows a sequence of the electron spectra as the electron
acceleration progresses.

Let us consider the radio emission produced by accelerated electrons
with the spectrum (\ref{el_spectrum_st_acc}) at the acceleration
region. We note that  gyrosynchrotron (GS) emission by
nonrelativistic and weakly relativistic electrons available during
an initial phase of the acceleration modeled by
Eq.~(\ref{el_spectrum_st_acc}) is inefficient; the flux of the GS
emission remains typically very small until sufficient number of
electrons is accelerated up to a few hundred keV\footnote{We note
that in case of big flares, large numbers of GS-producing electrons
can already be generated during a preflare phase
\citep{Asai_etal_2006}. In such cases we have in mind an even
earlier stage of acceleration \citep[e.g.,][]{Battaglia_etal_2009},
when the 100 keV electrons are not yet numerous.} \citep{BBG}.
However, along with the regular magnetic field, there is a spectrum
of turbulent waves (those accelerating the electrons) at the
acceleration site. The nonthermal electrons, interacting with those
random waves, experience spatial diffusion and so produce so called
Diffusive Synchrotron Radiation \citep[DSR,][]{Fl_2006a}, which we
calculate here.

Neglecting for simplicity the plasma gyrotropy we can take the
refractive index of the radio waves in the form
\begin{eqnarray}
\label{ref_ind} n_\sigma =\sqrt{\varepsilon}, \qquad
\varepsilon&=&1-\frac{\omega_{pe}^2}{\omega^2}, \qquad
\omega_{pe}=\sqrt{\frac{4\pi e^2n_e}{m}}.
\end{eqnarray}
Then,  spectral and angular distribution of the energy radiated by a
relativistic charged particle with a given Fourier transformed
acceleration $\mathbf{w}_{\omega'}$ during time $T$ of the particle
motion in an external field is given by \citep{LL_1971}
\begin{equation}
\label{cal_E_w_rel}
  {\cal E}_{{\bf n},\omega}=\sqrt{\varepsilon}\frac{Q^2}{c^3}  \left(\frac{\omega}{\omega
  '}\right)^4 \left|
  \left[{\bf n}\left[\left({\bf n}-\frac{\bf
  v}{c}\right){\bf w}_{\omega '}\right]\right] \right|^2
  ,
\end{equation}
where
\begin{equation}
\label{omega_prime}
  \omega '= \omega  \left(1-\frac{{\bf
  nv}}{c}\sqrt{\varepsilon(\omega)}\right).
\end{equation}

In the nonrelativistic case $v/c \ll 1$
($\gamma\equiv\sqrt{1-v^2/c^2} \approx 1$) and $\omega' \approx
\omega$, Eq. (\ref{cal_E_w_rel}) reduces to

\begin{equation}
\label{cal_E_w}
  {\cal E}_{{\bf n},\omega}=\sqrt{\varepsilon}\frac{Q^2}{c^3}
  \left|
  \left[{\bf n}\times{\bf w}_{\omega }\right] \right|^2
  ,
\end{equation}
where  $Q$ is the particle charge and ${\bf n}$ is the unit wave
vector of the radiation. Eq. (\ref{cal_E_w}) shows that the
radiation in a given direction ${\bf n}$ is defined by the
acceleration component $\left|
  {\bf w}_{\omega\bot } \right|^2=\left|
  \left[{\bf n}\times{\bf w}_{\omega }\right] \right|^2$
transverse to ${\bf n}$. Similarly to the derivation in
ultrarelativistic case \citep{Fl_2006a}, the transverse component of
the acceleration can be expressed via temporal and spatial Fourier
transform of the external Lorentz force, $F^\alpha_{q_0, {\bf q}}$
\begin{equation}
\label{w_perp_3}
 \mid {\bf w}_{\omega\bot} \mid^2=\frac{(2\pi)^3}{M^2  V}
 \int  dq_0 d{\bf q}
  \delta(\omega-q_0+{\bf qv})(\delta_{\alpha\beta}-n_\alpha n_\beta)
  F^\alpha_{q_0, {\bf q}}F^{\beta *}_{q_0, {\bf q}},
\end{equation}
where $M$ is the mass of emitting particle and $V$ is the source
volume. 

For  electric component of the Lorenz force $\mathbf{F}=Q\mathbf{E}$
we have

\begin{equation}
\label{E_field}
 (\delta_{\alpha\beta}-n_\alpha n_\beta)
  F^\alpha_{q_0, {\bf q}}F^{\beta *}_{q_0, {\bf q}}=
  Q^2(\delta_{\alpha\beta}-n_\alpha n_\beta)
  E^\alpha_{q_0, {\bf q}}E^{\beta *}_{q_0, {\bf q}},
\end{equation}
where
\begin{equation}
\label{E_corr}
   E^\alpha_{q_0, {\bf q}}E^{\beta *}_{q_0, {\bf q}}
   = \frac{TV}{(2\pi)^4} K_{\alpha
\beta}(q_0,{\bf q}),
\end{equation}
$K_{\alpha \beta}(q_0,{\bf q})$ is the correlation tensor of the
random electric field, such as $\int dq_0 d{\bf q} K_{\alpha
\alpha}(q_0,{\bf q})=\left<E_{st}^2\right>$ \citep{Toptygin_1985}.

For  magnetic component of the Lorenz force the corresponding
expression is different
\begin{equation}
\label{B_field}
\begin{array}{l}
 (\delta_{\alpha\beta}-n_\alpha n_\beta)
  F^\alpha_{q_0, {\bf q}}F^{\beta *}_{q_0, {\bf q}}=
  \frac{Q^2}{c^2}\left(v^2 \delta_{\alpha\beta} -
  v_\alpha v_\beta -[\mathbf{n}\times \mathbf{v}]_\alpha [\mathbf{n}\times \mathbf{v}]_\beta
  \right)
  B^\alpha_{q_0, {\bf q}}B^{\beta *}_{q_0, {\bf q}}=\\\\
  Q^2\frac{v^2}{c^2}\left(n_\alpha n_\beta+\frac{({\bf nv})^2}{v^2}\delta_{\alpha\beta}-
  ({\bf nv})\frac{v_\alpha n_\beta+n_\alpha v_\beta}{v^2}\right)
  B^\alpha_{q_0, {\bf q}}B^{\beta *}_{q_0, {\bf q}}.
\end{array}
\end{equation}
Similarly to Eq. (\ref{E_corr}) we have

\begin{equation}
\label{B_corr}
   B^\alpha_{q_0, {\bf q}}B^{\beta *}_{q_0, {\bf q}}
   = \frac{TV}{(2\pi)^4} K_{\alpha
\beta}(q_0,{\bf q}),
\end{equation}
where $K_{\alpha \beta}(q_0,{\bf q})$ is the correlation tensor of
the random magnetic field, such as $\int dq_0 d{\bf q} K_{\alpha
\alpha}(q_0,{\bf q})=\left<B_{st}^2\right>$. Thus, the DSR
intensity, $I_{\textbf{n},\omega}={\cal E}_{{\bf n},\omega}/T$,  of
a nonrelativistic particle in the presence of random magnetic field
is
\begin{equation}
 \label{I_DSR_nw_gen}
 I_{\textbf{n},\omega} =\sqrt{\varepsilon} \frac{Q^4 v^2}{2\pi M^2 c^5}
 \int dq_0 d{\bf q} \delta(\omega-q_0+\textbf{qv})
 \left(n_\alpha n_\beta+\frac{({\bf nv})^2}{v^2}\delta_{\alpha\beta}-
  ({\bf nv})\frac{v_\alpha n_\beta+n_\alpha v_\beta}{v^2}\right)K_{\alpha
\beta}(q_0,{\bf q}).
\end{equation}
This expression is valid for arbitrary spectrum of  magnetic
turbulence including anisotropic distributions.

We consider here the DSR produced by accelerated nonrelativistic
electrons interacting with the MHD turbulence. In MHD waves $E \sim
(v_a/c) B$, where $v_a$ is the Alfv\'en speed, therefore the
magnetic part of the Lorenz force is larger than the electric part
for all electrons with $v>v_a$. Assuming this condition to be
fulfilled, we calculate only the DSR related to the magnetic field
of the MHD turbulence; inclusion of electric field effect will
further increase the DSR intensity.

Since we are interested in overall spectral shapes and flux level of
the DSR, rather than   model-dependent details of the emission, we
consider here the simplest case of the isotropic MHD turbulence:

\begin{equation}
\label{B_corr_iso}
K_{\alpha\beta}=\frac{1}{2}\left({\delta}_{\alpha\beta}-\frac{q_{\alpha}q_{\beta}}
{q^{2}}\right)K(\textbf{q})\delta(q_{0}-q_{0}(\textbf{q})).
\end{equation}
As we assumed $v>v_a$, i.e., the electrons move faster than the
waves, we can adopt the MHD turbulence to be quasi static,
$q_{0}(\textbf{q})=0$. When the MHD turbulence is isotropic, the
accelerated electrons are isotropic as well, and so the radiation
produced is also isotropic. Thus, we consider further the radiation
produced into the full solid angle

\begin{equation}
\label{I_w_def}
 I_{\omega} = \int I_{\textbf{n},\omega}d\Omega =
 \sqrt{\varepsilon} \frac{8 Q^2}{3\pi c}\cdot q(\omega) 
 ,
\end{equation}
where, like in the ultrarelativistic case \citep{Fl_2006a}, we
introduce the scattering rate of the nonrelativistic particle by MHD
turbulence $q(\omega)$:

\begin{equation}
\label{q_w_iso} q(\omega)=\frac{\pi}{4}\left(\frac{Q}{Mc}\right)^{2}
\frac{v^2}{c^2}\int
K(\textbf{q})\delta(\omega+\textbf{qv})\,{d\textbf{q}}.
\end{equation}

To proceed further we have to specify the shape of the turbulence
spectrum $K(\textbf{q})$; we adopt a single power-law down to the
smallest (resonant to  thermal electrons) scales:

\begin{equation}
\label{K_pow_law} K(\textbf{q})=\frac{A_{\nu}}{q^{\nu+2}}\qquad
A_\nu=\frac{\nu-1}{4\pi}k_0^{\nu-1}\langle B_{st}^2\rangle,
\end{equation}
where $k_0=2\pi/L_0$ with $L_0$ the largest turbulence scale,
$\langle B_{st}^2\rangle$ is the mean square of the turbulent
magnetic field, and $\nu$ is the turbulence spectral index.

Then, substituting (\ref{K_pow_law}) into (\ref{q_w_iso}),
integrating over $d\textbf{q}$,

\begin{equation}
\begin{array}l
\int\,{d\textbf{q}}K(\textbf{q})\delta(\omega+\textbf{qv})
=2\pi\int\,{d}\cos\theta\cdot\,{d}q\frac{A_\nu}{q^\nu}\delta(\omega+qv\cos\theta)=
\frac{2\pi
A_\nu}{v}\int\limits_{\frac{\omega}{v}}^{\frac{\omega_{pe}}{v_{pe}}}\frac{\,d
q}{q^{\nu+1}} =\\
 \frac{2\pi}{\nu}\frac{A_\nu}{v}\left(\frac{v}{\omega}\right)^\nu
 \left(1-\left(\frac{\omega v_{pe}}{\omega_{pe} v}\right)^\nu\right)
 \Theta\left(\frac{\omega_{pe}}{v_{pe}}-\frac{\omega}{v}\right)
\end{array}
\end{equation}
where
\begin{eqnarray}
 \label{v_therm}
v_{pe}&=&6.74\times10^5\sqrt{T_e}
\end{eqnarray}
is the thermal velocity of the plasma electrons, $\Theta(x)$ is the
step function, and using the electron charge $e$ and mass $m$ for
$Q$ and $M$, we find
\begin{equation}
\label{q_w_PLW}
 q(\omega)=\frac{\pi^2A_\nu}{2\nu}
 \frac{e^2 v}{m^2c^4}\left(\frac{v}{\omega}\right)^\nu
 \left(1-\left(\frac{\omega v_{pe}}{\omega_{pe} v}\right)^\nu\right)
 \Theta\left(\frac{\omega_{pe}}{v_{pe}}-\frac{\omega}{v}\right),
\end{equation}
so the DSR spectrum produced by accelerated electrons reads
\begin{eqnarray}
\label{DSR_w_eps}  I_{\omega}&=& \frac{8e^2}{3\pi
c}\sqrt{\varepsilon}\cdot q(\omega).
\end{eqnarray}



Now we calculate the DSR power from $N$ electrons with the spectrum
described by Eq. (\ref{el_spectrum_st_acc})
\begin{equation}
\label{DSR_ensemble} P_\omega=\int\limits_{E_0}^\infty I_\omega
N(E)\,dE.
\end{equation}
In fact, we are interested in the radio flux observed at the Earth.
To transform this radiation power into the flux observed at the
Earth, we change the variable $\omega=2\pi f$, so that $I_f=2\pi
I_\omega$. Then, the flux is
\begin{equation}
\label{DSR_flux} F_f=\frac{2\pi P_\omega V}{4\pi
R_{au}^2}=\frac{P_\omega L^3}{2R_{au}^2}\cdot10^{19} \quad {\rm
sfu},
\end{equation}
where $R_{au}=1$~au$=1.49\times10^{13}$~cm is the distance from the
Earth to the Sun.


To evaluate the DSR from the acceleration region of a solar flare,
we adopt some typically assumed parameters of the acceleration site
as follows: (a) the size of the site $L\sim 10^8$ cm; (b) the
thermal electron number density $n_e\sim 10^{10}$ cm$^{-3}$; (c) the
electron temperature $T_e\sim 10^6$ K; (d) the energy density of the
magnetic turbulence $W_{st}=\frac{\langle
B_{st}^2\rangle}{8\pi}\sim10^3$ erg/cm$^{3}$. Accordingly, the total
energy, $W_{tot}\sim W_{st}L^3 \sim 10^{27}$ ergs, corresponds to a
very modest solar flare. We assume that the power-law tail of the
accelerated electrons grows from $E_0=(4-6) kT_e$ with $n(>E_0)$
specified by matching condition (\ref{el_matching_st_acc}) and as
the acceleration has started, the power law index $\delta$ changes
from $8\sim3$ while the break energy $E_{br}$
increases from $50\sim500$ keV.

\begin{figure*}[!ht]
\begin{center}
\epsscale{1} 
 \plotone{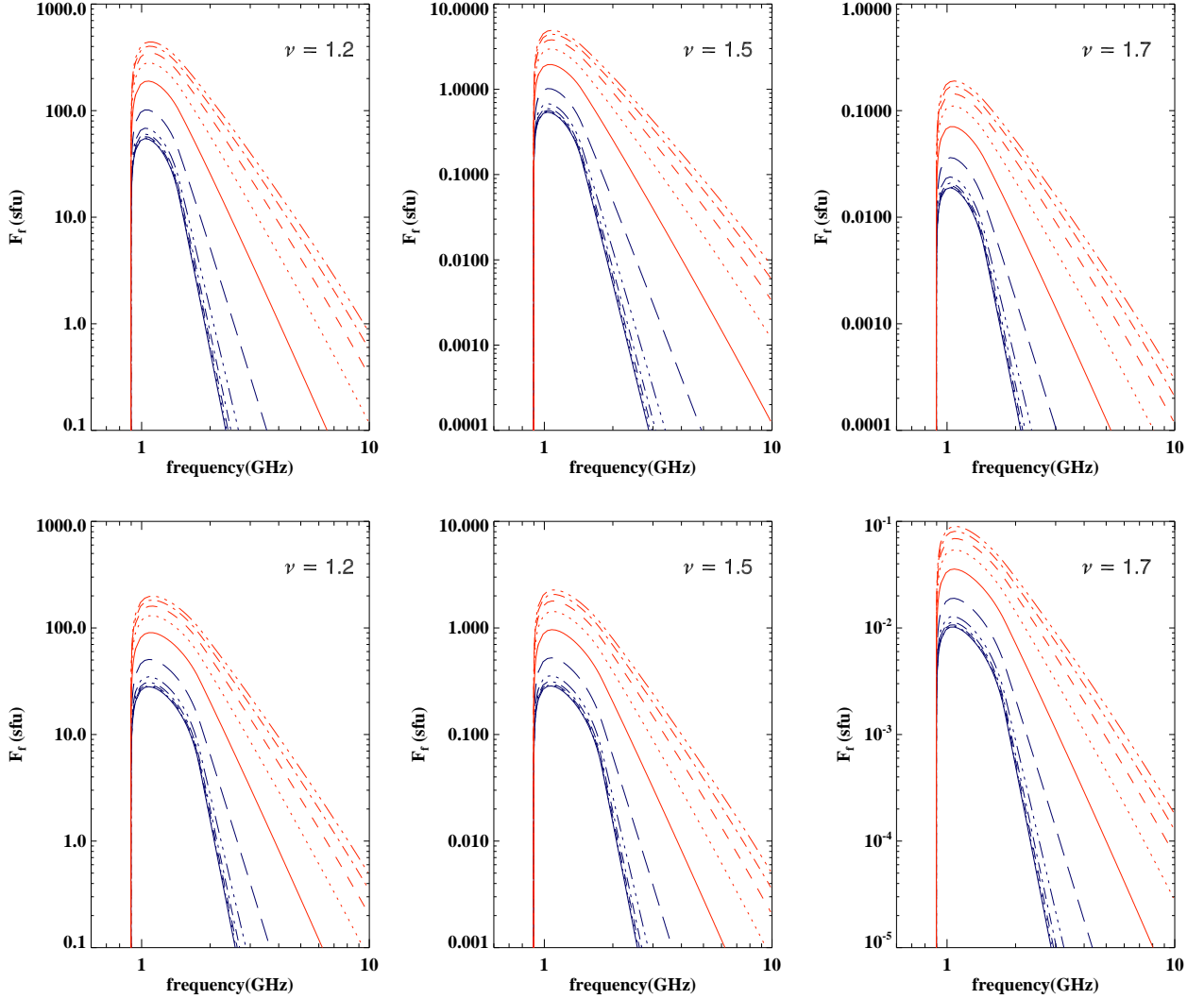}\caption{The DSR total flux density
spectra calculated for the observations from the Earth surface for
$E_0=4kT$ (upper panel) and $E_0=6kT$ (lower panel) for 11 different
$\delta$ from $7\sim3$ and three different $\nu$, indicated in the
panels. The blue curves indicate larger $\delta$, while the red ones
show smaller $\delta$.} \label{FIG03}
\end{center}
\end{figure*}

\begin{figure*}[t]
\centering
  \subfigure[]{\label{fig:edge-a}\includegraphics[width=3.3in]{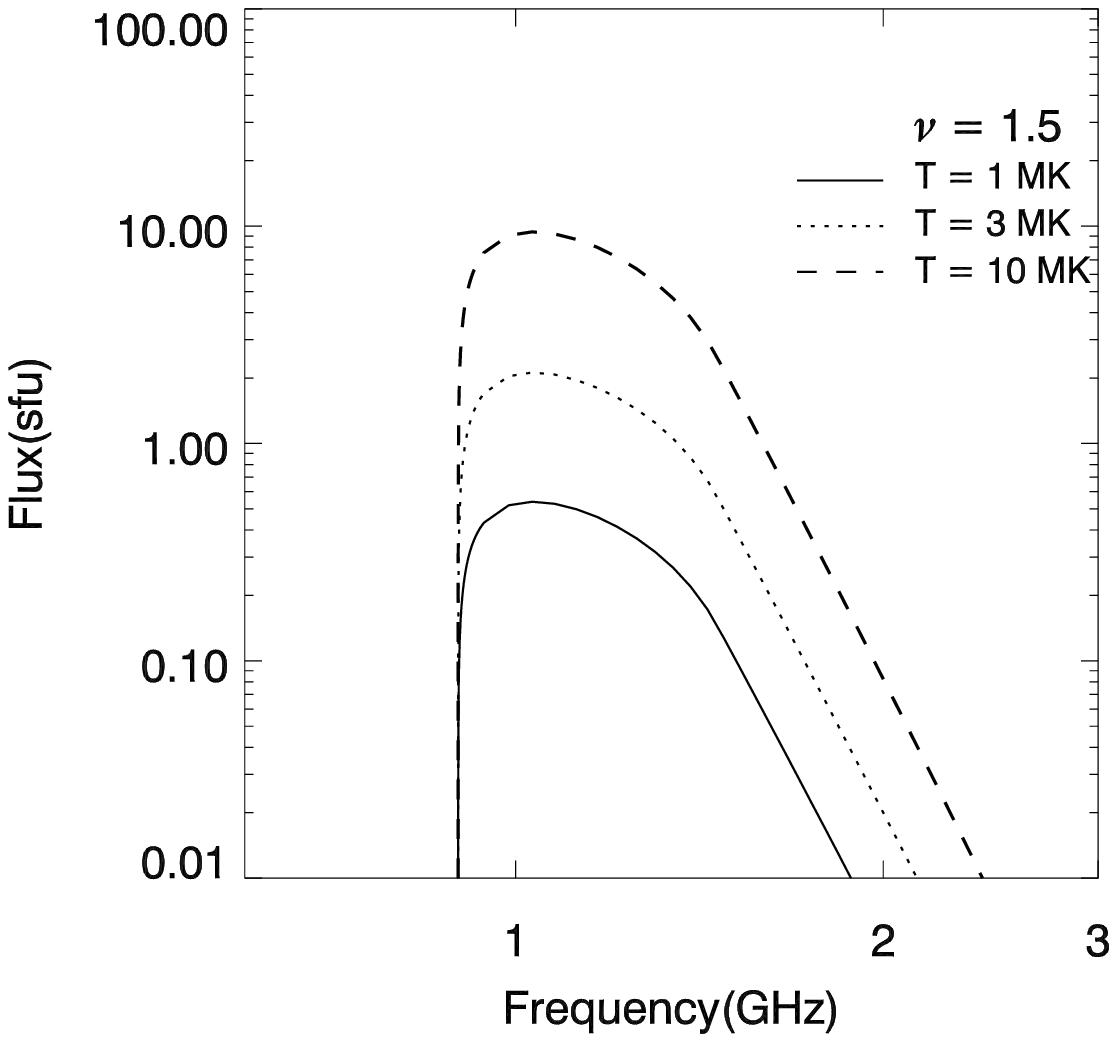}}
  \subfigure[]{\label{fig:edge-b}\includegraphics[width=3.3in]{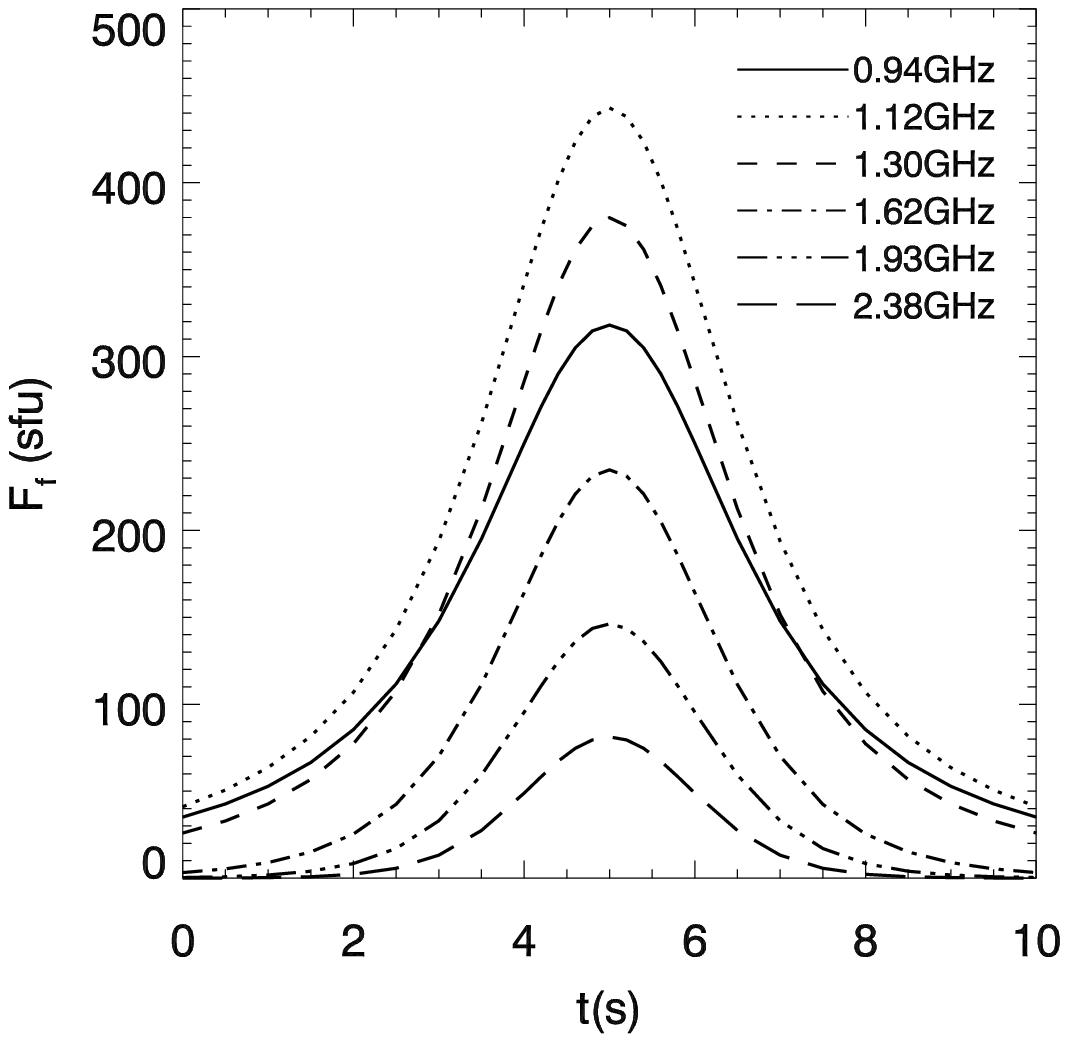}}\\
 \caption{(a)  DSR
spectra for  three sources with $\delta=7$, $E_0=4kT_e$, and
different electron temperatures,  $T_e= (1,~3,~10)\cdot 10^6$ K; (b)
DSR light curves for the case $E_0=4kT_e$, $\nu=1.2$,
$E_{br}=\left(E_{br}(t=0)+\frac{dE_{br}}{dt}\cdot t\right)$ keV,
where $E_{br}(t=0)=50$~keV, $\frac{dE_{br}}{dt}=45$~keV/s, and
soft-hard-soft evolution of the accelerated electron spectrum,
$\delta = \delta_{\max} + a\cdot (t-t_0)^2$, where $\delta_{\max}
=3$, $t_0=5$~s, and $a=0.2$~s$^{-2}$. The apparent symmetry of the
model light curves is provided by assumed symmetry of the spectral
index variation and adopted no variation of MHD spectrum; relaxing
any of these assumption will result in asymmetric light curves. All
the light curves peak almost simultaneously, so no
"cross-correlation" delay is expected, although onset of higher
frequency light curves is delayed, thus, the total duration of
higher frequency emission is shorter. } \label{FIG04}
\end{figure*}

Figure~\ref{FIG03} presents the sequence of  calculated DSR spectra
for 11 different $\delta$ values from $7\sim3$; the spectra are
calculated for three different $\nu$ values and for two different
$E_0$ values. The blue curves indicate larger $\delta$, while the
red ones show smaller $\delta$. Then, Figure~\ref{FIG04}(a) presents
the DSR spectra for three different temperature values, $T=
(1,3,10)\cdot 10^6 K$. In addition to spectrum shapes, light curves
of the radiation at different frequencies can be informative. To
estimate the light curve behavior we adopt a soft-hard-soft spectrum
evolution as follows from theory of spectrum evolution for the
stochastic acceleration \citep{Byk_Fl_2009}, and which is  typical
for impulsive flares, with the electron energy spectral index
$\delta(t)$ changing from  8 to 3 and then back to 8, while the
break energy $E_{br}$ increasing all the way from 50 keV to $\sim
1$~MeV. Figure~\ref{FIG04}(b) shows the corresponding model light
curves at a few frequencies around the spectrum peak. One can note
from the figure that higher frequency light curves have a somewhat
shorter duration, although peaking at the same time; so no
appreciable time delay between the light curves is expected.

We note that the DSR spectra are very narrowband, much narrower that
typical gyrosynchrotron spectra. The high frequency slope of the DSR
spectrum can easily be evaluated from Eqns. (\ref{DSR_flux}),
(\ref{DSR_ensemble}), (\ref{q_w_PLW}), and (\ref{DSR_w_eps}), $F_f
\propto f^{3-2\delta}$. Thus, the DSR high frequency spectral index
varies from 11 to 3 as the spectral index of accelerated electrons
changes from 7 to 3, while the GS spectral index would vary from 5
to 1 for the same range of $\delta$ variation. The peak flux of the
DSR is highly sensitive to the turbulence spectral index (specified
eventually by the MHD cascading law), while less sensitive to the
plasma temperature and electron spectral index. The peak flux can be
very large (up to a few hundred sfu), which makes it easily
observable even by  full sun radio instruments. If so, the
corresponding radio emission must have been widely observed by
available radio spectrometers working in the decimetric and/or
microwave range. Indeed, there is a class of radio bursts with the
properties resembling the DSR properties described here---it is the
class of narrowband decimetric and microwave continuum bursts
(including type IVdm), which, we suggest, may contain
burst-candidates to the radio emission from the regions of
stochastic acceleration in  solar flares. Although this
interpretation is tempting, spatially resolved radio observations of
the DSR will be needed to confirm it, to locate the region of
stochastic acceleration, and study it in detail. Another plausible
candidate for radio emission from stochastic acceleration episodes
is so-called transient brightenings, whose radio spectra are often
narrowband \citep{Gary_etal_1997}.

\section{Gyrosynchrotron Radio Emission from a Collapsing Trap}

Let us consider another model, a collapsing magnetic trap, which can
efficiently accelerate charged particles. Unlike the stochastic
acceleration models, no turbulence spectrum is essential to
accelerate particles in the collapsing trap model; however, some
spectrum of 'pre-accelerated' particles is needed, otherwise, the
collapse of the trap will only give rise to plasma heating without
nonthermal particle generation.

Therefore, we assume that just before collapsing the trap contained
both thermal plasma and nonthermal electron population with a
power-law spectrum. To be specific, we adopt  the initial conditions
as follows: (a) the magnetic field strength $B_0=30$ G; (b) the
minimum and the maximum energy of the power-law spectrum $E_{\rm
min}=0.01$ MeV, $E_{\rm max}=1$ MeV; (c) the thermal electron
density $n_{th}=10^9$ cm$^{-3}$ and the non-thermal electron density
$n_{rl}=10^7$ cm$^{-3}$; (d) the source size $L_0=10"$.

\begin{figure}[!ht]
 \centering
  \subfigure[]{\label{fig:edge-a}\includegraphics[width=3.3in]{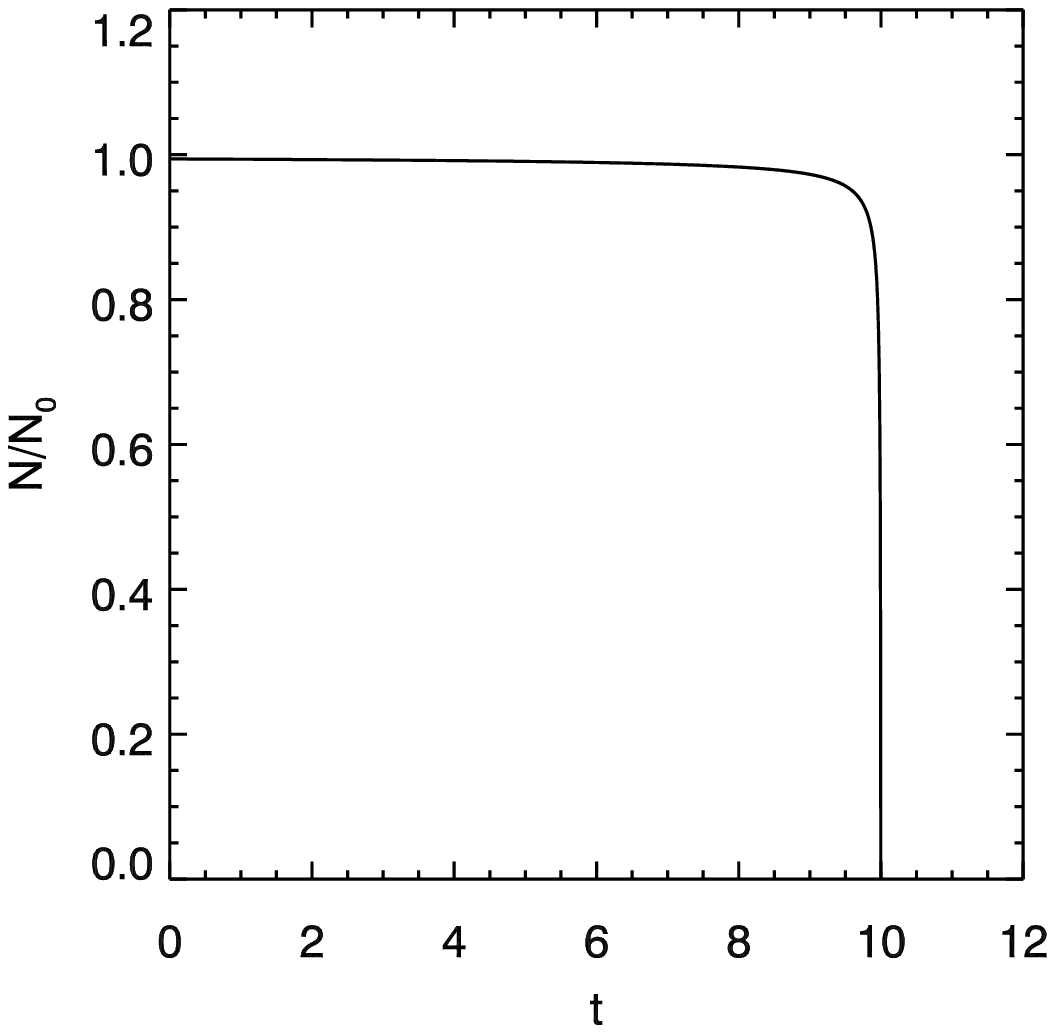}}
  \subfigure[]{\label{fig:edge-b}\includegraphics[width=3.3in]{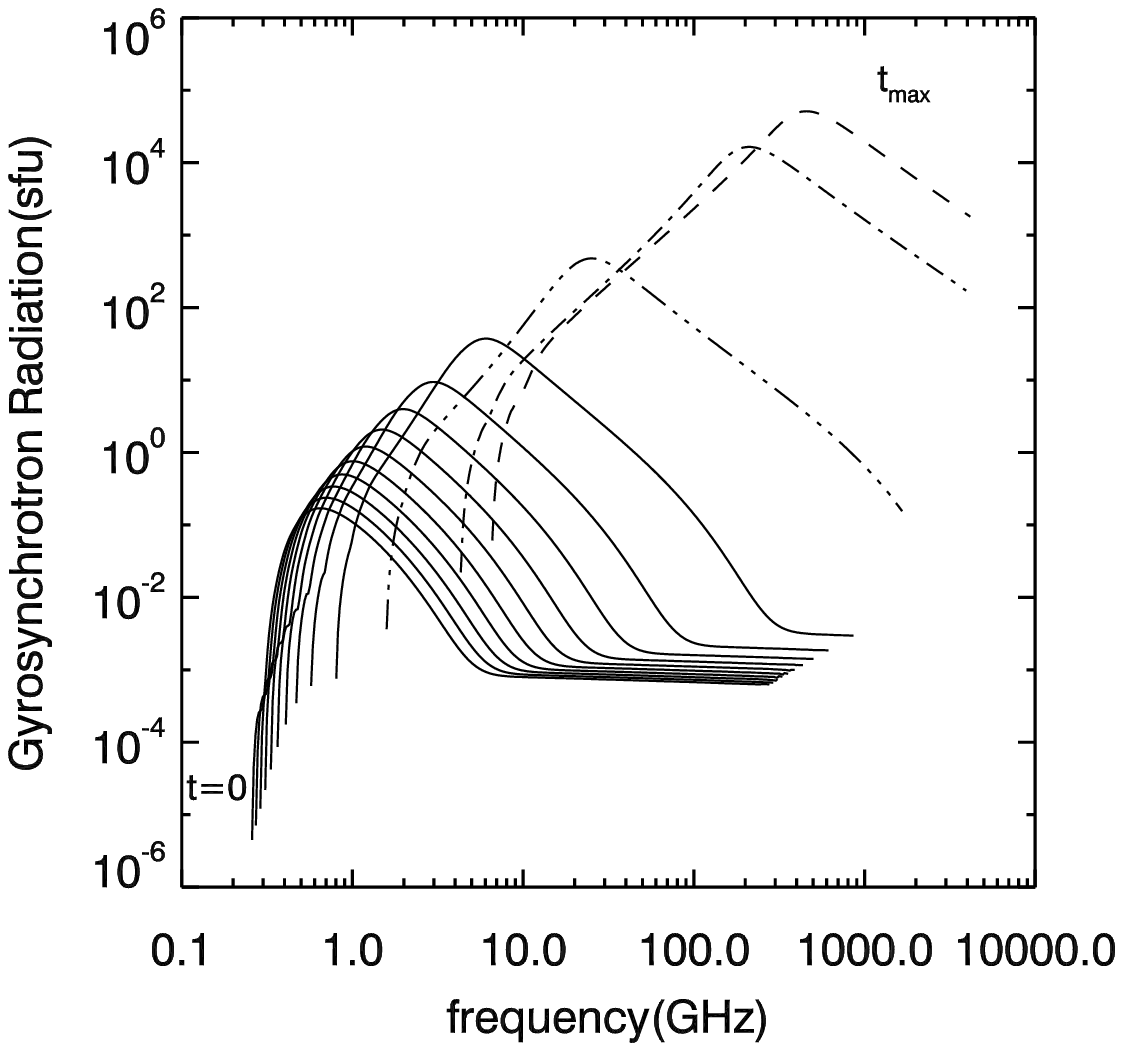}}\\
  \caption{Left: The number of electrons in the trap according to the
\citet{Bogachev_Somov_2005} model. Right: GS spectra in frequency
range $0.1\sim10^4$~GHz; the solid spectra are plotted for $t=$ 0,
1, ..., 9~s; the dashed-doted lines correspond to the end of
collapse at $t=$ 9.75, 9.975, 9.9975~s. In fact, we take into
account both GS and free-free contributions to the radiation
intensity, however, the free-free component is inessential during
the loop collapse: it is only noticeable at high frequency flat
regions of the early spectra.} \label{FIG05}
\end{figure}

During the trap contraction, the number of accelerated electrons
evolves. For our modeling we adopt a solution obtained by
\citet{Bogachev_Somov_2005}, see Figure~\ref{FIG05}, which takes
into account the betatron and Fermi acceleration and the particle
escape from the trap via the loss cone:

\begin{equation}
N=N_0\frac{l\sqrt{b_m-b}}{\sqrt{1+(b_m-1)l^2}}
\end{equation}
where
\begin{eqnarray}
b=b(t)&=&B(t)/B_0  \\
l=l(t)&=&L(t)/L_0
\end{eqnarray}
so $b(t)$ changes from $b(0)=1$ to $b_m=B_m/B_0$, $B_m$ is the
largest magnetic field value at the end of the trap collapse, and
$l(t)$ deceases from $l(0)=1$ to a very low value, say, $l(t_{\rm
max})=0.1$. For the sake of simplicity we assume a self-similar
contraction of the collapsing magnetic trap. In this case, evolution
of all parameters of the trap is uniquely defined by their initial
values and the dimensionless source scale $l(t)$. Thus, for any
given contraction law, $l(t)$, we can easily calculate the
corresponding time history of all other relevant source parameters,
as the magnetic field, the thermal electron number density, the
source volume and the projected area, and the evolution of the
nonthermal electron spectrum \citep{Bogachev_Somov_2005,
Bogachev_Somov_2007}. For our modeling we adopt that the trap volume
$V$ linearly decreases with time during the trap contraction from
$10"^3$ to $1"^3$; we adopt 10~s for the trap collapse time, which
is a few Alfven times ($\tau_a\sim L/v_a$) for the trap parameters
used.

Thus, we can straightforwardly calculate the GS spectra at different
time frames and the radio light curves at different frequencies
within the adopted collapsing trap model. Figure~\ref{FIG05}
displays the GS spectra at different moments of the trap
contraction. In agreement with a statement made in the previous
section, at initial phase of acceleration the GS flux is  small
(less than 1 sfu), which can only be recorded by high sensitivity
spatially resolved observations. However, during the trap
contraction the magnetic field increases and the fast electrons are
accelerated, which all together lead to a significant increase of
the peak flux and the peak frequency of the radio emission produced
at the acceleration site; thus the radio emission becomes easily
detectable by available radio instrument soon after the trap starts
to contract.

\begin{figure}[!ht]
 \centering
  \subfigure[]{\label{fig:edge-a}\includegraphics[width=3.3in]{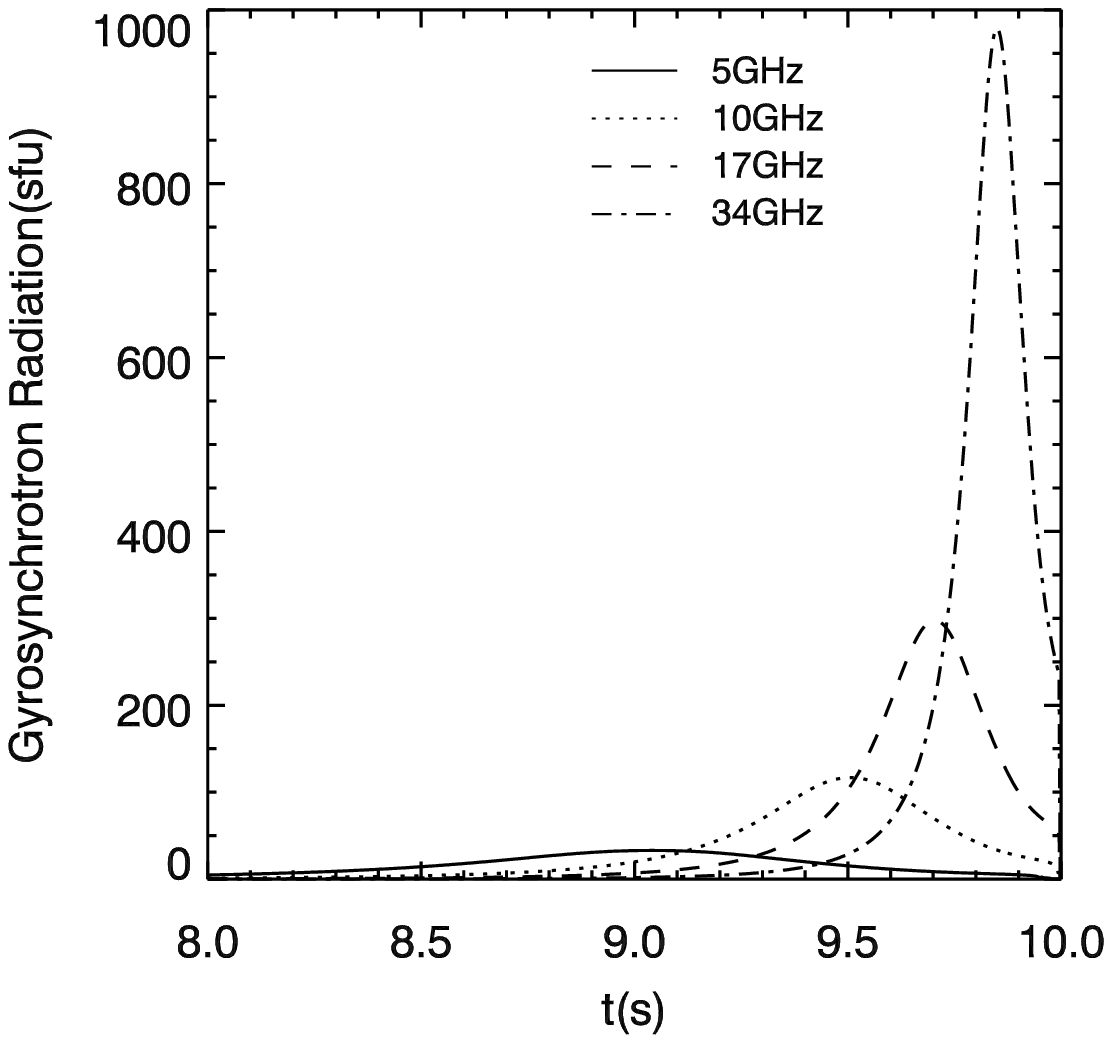}} 
  \subfigure[]{\label{fig:edge-b}\includegraphics[width=3.3in]{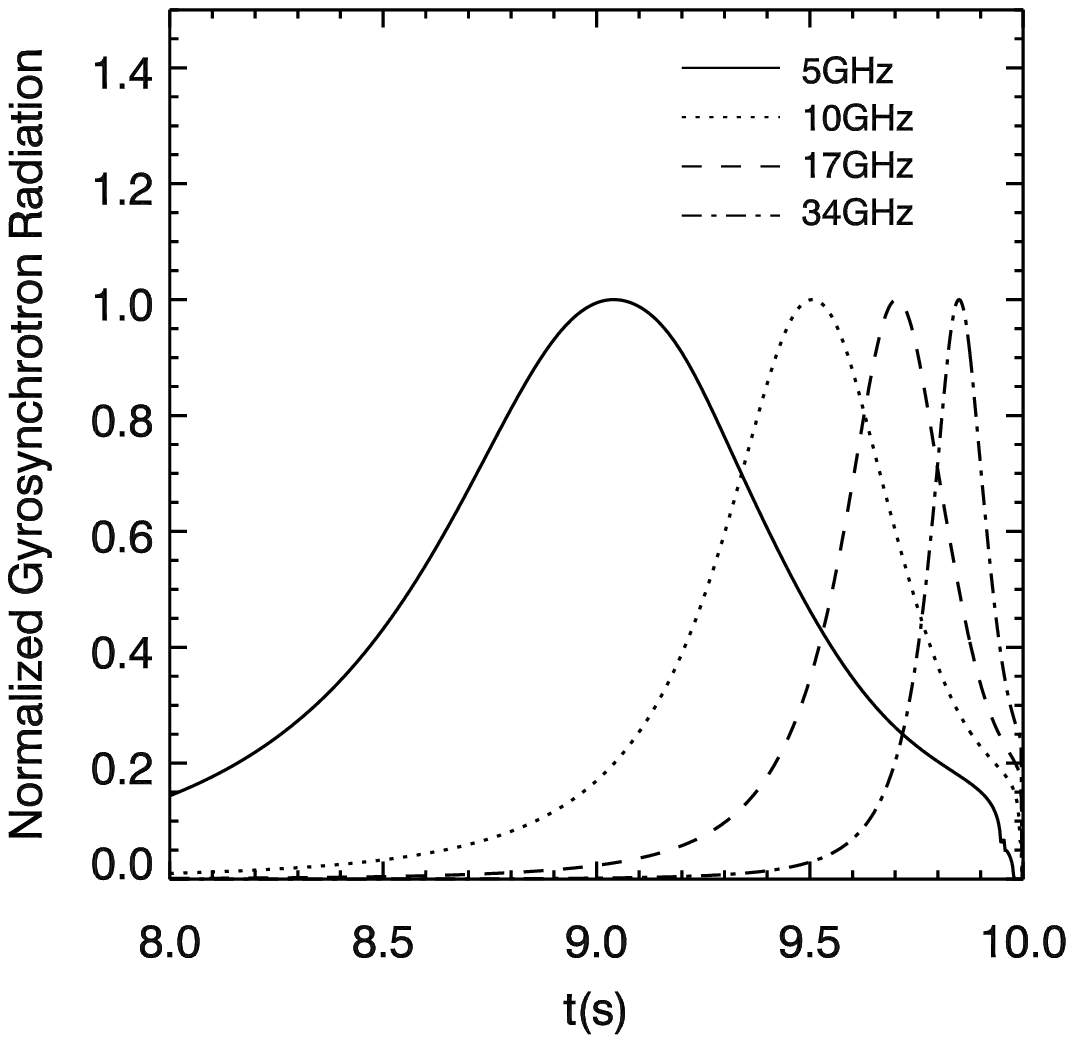}}\\
  \caption{The light curves (a) and the normalized light curves (b)
  of the GS emission from a collapsing trap at a
number of fixed frequencies  5, 10, 17, 34 GHz. The initial trap
parameters and the contraction law are described in the text. A
significant time delay between the light curves, comparable to
duration of the light curves, is evident from the Figure.}
 \label{FIG07}
\end{figure}

Figure~\ref{FIG07} presents the light curves of the emission at a
number of fixed frequencies,  5, 10, 17, 34 GHz. Within the adopted
model the peak flux increases with frequency, see
Figures~\ref{FIG05},~\ref{FIG07}; in fact, this increase may become
less pronounced if the coulomb losses in the collapsing trap are
taken into account \citep{Bogachev_Somov_2009}. A distinctive
feature of the light curves, contrasting to that of DSR produced
from the stochastic acceleration sites, is a noticeable time delay:
the higher frequency light curves are delayed relative to lower
frequency light curves; this time  delay will be present even when
the coulomb losses \citep{Bogachev_Somov_2009} are included. A time
delay in the sense predicted by our modeling is frequently observed
in solar flares, in particular, in those with quasiperiodic
pulsations \citep{Fl_etal_2008}. Observationally, however, the GS
emission from a collapsing trap can be contaminated by GS emission
from trapped electrons produced by previous acceleration episodes,
so unambiguous detection of the GS emission from a collapsing trap
itself requires additional accurate analysis to separate the
contributions, which as yet has not been performed.


\section{Discussion}

There are many models in which  electrons can be  accelerated to
nonthermal energies. Some mechanisms  accelerate a tiny fraction of
the electrons, which can only be observed via coherent radio
emissions (e.g., type III bursts produced by electron beams, or
accompanying metric spikes), others produce more powerful
acceleration, sufficient to generate observable incoherent radio
emission from either the acceleration site itself of from a remote
'radiation site'.

The idea of using radio observations to probe energy
release/acceleration regions in flares has been around for awhile
\citep[e.g.,][]{BBG}, however, the studies focused mainly on
\emph{coherent} decimeter radio bursts. For example,
\cite{Benz_1986} argued that decimeter narrowband millisecond radio
spike clusters can be a signature of electron acceleration in
flares, and, if so, the flare energy release must have been highly
fragmented with each spike indicating a single energy
release/acceleration episode. However, it has been found
\citep{Aschwanden_Guedel_1992} that the radio spikes are frequently
delayed compared with associated hard X-ray emission, implying the
spikes are a secondary phenomenon associated with flares. Moreover,
spatially resolved observations \citep{Benz_etal_2002,
Battaglia_Benz_2009} show that the spike sources are typically far
away from main flare locations. Even though higher frequency
microwave radio spikes \citep{spikes, Rozh_etal_2008} can be
produced at or around the main flare location \citep{Gary_2009}, it
seems doubtful that the coherent radio burst originate  from
elementary acceleration episodes   \citep{Fl_Meln_1998, spikes,
Rozh_etal_2008, Battaglia_Benz_2009}.

In contrast, in this Letter we have calculated \emph{incoherent}
radio emission from the acceleration region of a solar flare within
two distinct acceleration models---stochastic acceleration by
cascading MHD turbulence and regular (betatron and Fermi)
acceleration in a collapsing trap. We have demonstrated that  the
radio emissions produced within these two competing acceleration
models are distinctly different, which potentially allows
distinguishing between them by the radio observations. In
particular, we have found that the stochastic acceleration process
is accompanied by a very narrowband  DSR continuum radio emission,
whose predicted properties are generally consistent with  observed
properties of narrowband microwave or decimetric  (type IVdm)
continuum bursts, thus, we suggest that some of those bursts can be
produced from the sites of stochastic acceleration.
%

\acknowledgments  This work was supported in part by NSF grants
AST-0607544, ATM-0707319, and ATM-0745744,  and NASA grant
NNG06GJ40G, NNX0-7AH78G, and NNX0-8AQ90G to New Jersey Institute of
Technology, and by the Russian Foundation for Basic Research, grants
08-02-92228, 09-02-00226, and 09-02-00624. We have made use of
NASA's Astrophysics Data System Abstract Service.

%

\begin{thebibliography}{38}
\expandafter\ifx\csname
natexlab\endcsname\relax\def\natexlab#1{#1}\fi

\bibitem[{{Asai} {et~al.}(2006){Asai}, {Nakajima}, {Shimojo}, {White},
  {Hudson}, \& {Lin}}]{Asai_etal_2006}
{Asai}, A., {Nakajima}, H., {Shimojo}, M., {White}, S.~M., {Hudson},
H.~S., \&
  {Lin}, R.~P. 2006, \pasj, 58, L1

\bibitem[{{Aschwanden}(2002)}]{Aschw_2002}
{Aschwanden}, M.~J. 2002, {Particle Acceleration and Kinematics in
Solar
  Flares} (Particle Acceleration and Kinematics in Solar Flares, A Synthesis of
  Recent Observations and Theoretical Concepts, by Markus J.~Aschwanden,
  Lockheed Martin, Advanced technology Center, palo Alto, California,
  U.S.A.~Reprinted from SPACE SCIENCE REVIEWS, Volume 101, Nos.~1-2 Kluwer
  Academic Publishers, Dordrecht)

\bibitem[{{Aschwanden} \& {G\"{u}del}(1992)}]{Aschwanden_Guedel_1992}
{Aschwanden}, M.~J. \& {G\"{u}del}, M. 1992, \apj, 401, 736

\bibitem[{{Bastian} {et~al.}(1998){Bastian}, {Benz}, \& {Gary}}]{BBG}
{Bastian}, T.~S., {Benz}, A.~O., \& {Gary}, D.~E. 1998, \araa, 36,
131

\bibitem[{{Battaglia} \& {Benz}(2009)}]{Battaglia_Benz_2009}
{Battaglia}, M. \& {Benz}, A.~O. 2009, \aap, 499, L33

\bibitem[{{Battaglia} {et~al.}(2009){Battaglia}, {Fletcher}, \&
  {Benz}}]{Battaglia_etal_2009}
{Battaglia}, M., {Fletcher}, L., \& {Benz}, A.~O. 2009, \aap, 498,
891

\bibitem[{{Benz}(1986)}]{Benz_1986}
{Benz}, A.~O. 1986, \solphys, 104, 99

\bibitem[{{Benz} {et~al.}(2002){Benz}, {Saint-Hilaire}, \&
  {Vilmer}}]{Benz_etal_2002}
{Benz}, A.~O., {Saint-Hilaire}, P., \& {Vilmer}, N. 2002, \aap, 383,
678

\bibitem[{{Bogachev} \& {Somov}(2005)}]{Bogachev_Somov_2005}
{Bogachev}, S.~A. \& {Somov}, B.~V. 2005, Astronomy Letters, 31, 537

\bibitem[{{Bogachev} \& {Somov}(2007)}]{Bogachev_Somov_2007}
---. 2007, Astronomy Letters, 33, 54

\bibitem[{{Bogachev} \& {Somov}(2009)}]{Bogachev_Somov_2009}
---. 2009, Astronomy Letters, 35, 57

\bibitem[{{Bykov} \& {Fleishman}(2009)}]{Byk_Fl_2009}
{Bykov}, A.~M. \& {Fleishman}, G.~D. 2009, \apjl, 692, L45

\bibitem[{{Fleishman}(2006)}]{Fl_2006a}
{Fleishman}, G.~D. 2006, \apj, 638, 348

\bibitem[{{Fleishman} {et~al.}(2008){Fleishman}, {Bastian}, \&
  {Gary}}]{Fl_etal_2008}
{Fleishman}, G.~D., {Bastian}, T.~S., \& {Gary}, D.~E. 2008, \apj,
684, 1433

\bibitem[{{Fleishman} {et~al.}(2003){Fleishman}, {Gary}, \& {Nita}}]{spikes}
{Fleishman}, G.~D., {Gary}, D.~E., \& {Nita}, G.~M. 2003, \apj, 593,
571

\bibitem[{{Fleishman} \& {Melnikov}(1998)}]{Fl_Meln_1998}
{Fleishman}, G.~D. \& {Melnikov}, V.~F. 1998, Uspekhi Fizicheskikh
Nauk, 41,
  1157

\bibitem[{{Gary} \& {Naqvi}(2009)}]{Gary_2009}
{Gary}, D.~E. \& {Naqvi}, M. 2009,  AAS Bull., 41, 851

\bibitem[{{Gary} {et~al.}(1997){Gary}, {Hartl}, \& {Shimizu}}]{Gary_etal_1997}
{Gary}, D.~E., {Hartl}, M.~D., \& {Shimizu}, T. 1997, \apj, 477, 958

\bibitem[{{Hamilton} \& {Petrosian}(1992)}]{Petrosian92}
{Hamilton}, R.~J. \& {Petrosian}, V. 1992, \apj, 398, 350

\bibitem[{{Holman}(1985)}]{Holman85}
{Holman}, G.~D. 1985, \apj, 293, 584

\bibitem[{{Holman} \& {Benka}(1992)}]{HolmanBenka}
{Holman}, G.~D. \& {Benka}, S.~G. 1992, \apjl, 400, L79

\bibitem[{{Karlick{\'y}} \& {Kosugi}(2004)}]{Karlicky_Kosugi_2004}
{Karlick{\'y}}, M. \& {Kosugi}, T. 2004, \aap, 419, 1159

\bibitem[{{Landau} \& {Lifshitz}(1971)}]{LL_1971}
{Landau}, L.~D. \& {Lifshitz}, E.~M. 1971, {The classical theory of
fields},
  ed. L.~D. {Landau} \& E.~M. {Lifshitz}

\bibitem[{{Litvinenko}(1996)}]{Litvinenko96}
{Litvinenko}, Y.~E. 1996, \apj, 462, 997

\bibitem[{{Litvinenko}(2000)}]{Litvinenko_2000}
---. 2000, \solphys, 194, 327

\bibitem[{{Litvinenko}(2003)}]{Litvinenko_2003a}
{Litvinenko}, Y.~E. 2003, in Lecture Notes in Physics, Berlin
Springer Verlag,
  Vol. 612, Energy Conversion and Particle Acceleration in the Solar Corona,
  ed. L.~{Klein}, 213--229

\bibitem[{{Miller}(1997)}]{Miller97}
{Miller}, J.~A. 1997, \apj, 491, 939

\bibitem[{{Miller} {et~al.}(1996){Miller}, {Larosa}, \& {Moore}}]{Miller96}
{Miller}, J.~A., {Larosa}, T.~N., \& {Moore}, R.~L. 1996, \apj, 461,
445

\bibitem[{{Nindos} {et~al.}(2008){Nindos}, {Aurass}, {Klein}, \&
  {Trottet}}]{Nindos_etal_2008}
{Nindos}, A., {Aurass}, H., {Klein}, K.-L., \& {Trottet}, G. 2008,
\solphys,
  253, 3

\bibitem[{{Park} {et~al.}(1997){Park}, {Petrosian}, \&
  {Schwartz}}]{Petrosian97}
{Park}, B.~T., {Petrosian}, V., \& {Schwartz}, R.~A. 1997, \apj,
489, 358

\bibitem[{{Petrosian} {et~al.}(1994){Petrosian}, {McTiernan}, \&
  {Marschhauser}}]{Petrosian94}
{Petrosian}, V., {McTiernan}, J.~M., \& {Marschhauser}, H. 1994,
\apj, 434, 747

\bibitem[{{Pryadko} \& {Petrosian}(1998)}]{Petrosian98}
{Pryadko}, J.~M. \& {Petrosian}, V. 1998, \apj, 495, 377

\bibitem[{{Rozhansky} {et~al.}(2008){Rozhansky}, {Fleishman}, \&
  {Huang}}]{Rozh_etal_2008}
{Rozhansky}, I.~V., {Fleishman}, G.~D., \& {Huang}, G.-L. 2008,
\apj, 681, 1688

\bibitem[{{Somov} \& {Bogachev}(2003)}]{Somov_Bogachev_2003}
{Somov}, B.~V. \& {Bogachev}, S.~A. 2003, Astronomy Letters, 29, 621

\bibitem[{{Somov} \& {Kosugi}(1997)}]{Somov_Kosugi_1997}
{Somov}, B.~V. \& {Kosugi}, T. 1997, \apj, 485, 859

\bibitem[{{Toptygin}(1985)}]{Toptygin_1985}
{Toptygin}, I.~N. 1985, {Cosmic rays in interplanetary magnetic
fields}, ed.
  I.~N. {Toptygin}

\bibitem[{{Tsuneta}(1985)}]{Tsuneta85}
{Tsuneta}, S. 1985, \apj, 290, 353

\bibitem[{{Vilmer} \& {MacKinnon}(2003)}]{Vilmer_MacKinnon_2003}
{Vilmer}, N. \& {MacKinnon}, A.~L. 2003, in Lecture Notes in
Physics, Berlin
  Springer Verlag, Vol. 612, Energy Conversion and Particle Acceleration in the
  Solar Corona, ed. L.~{Klein}, 127--160

\end{thebibliography}
%

%
%
%
%




\end{document}